# Phase engineering of 1T' and 1T $CrS_2$ and $Cr_2S_3$ by MOCVD

*Haoyu Bai[1], Gareth R.M Tainton[2,3], Mauro Och[1], Max Rimmer[2,3], Indrajit Maity[1], Filippo Mione[4], Khagesh Tanwar[4], Rongsheng Cai[3], Joseph Parker[3], Kho Zhiquan[3], Ercin C Duran[3], Evan Tillotson[2,3], Sam Sullivan-Allsop[2,3], Alexander Eggeman[3], Siyuan Deng[1], Dan Bromley[5], Alex Vanstone[5], Jack C. Gartside[5], Sandrine Heutz[1], Will R. Branford[5], Efrén Navarro-Moratalla[4], Johannes C. Lischner[1,6], Sarah J. Haigh[2,3], Cecilia Mattevi[1]**

[1] Department of Materials, Imperial College London, London, SW7 2AZ, United Kingdom.

[2] National Graphene Institute, University of Manchester, Oxford Road, Manchester M13 9PL, United Kingdom.

[3] Department of Materials, University of Manchester, Oxford Road, Manchester M13 9PL, United Kingdom.

[4] Instituto de Ciencia Molecular, Universitat de València, Paterna, 46980 Spain

[5] Department of Physics Imperial College London, London, SW7 2AZ, United Kingdom.

[6] Thomas Young Centre for Theory and Simulation of Materials, Imperial College London, London, SW7 2AZ, United Kingdom

*To whom correspondence should be addressed: c.mattevi@imperial.ac.uk

**Abstract**

Layered Cr-chalcogenides compounds offer a rich range of crystal phases with different magnetic properties some of which have been predicted only but not experimentally verified. Here we demonstrate a previously unreported crystal phase of $CrS_2$, namely the distorted octahedral (1T’) structure, synthesized via metal-organic chemical vapour deposition (MOCVD). We achieved the tuneable synthesis of either 1T’ or 1T phase $CrS_2$. The structure were identified using polarized Raman spectroscopy, density functional perturbation theory (DFPT), as well as scanning electron diffraction (4D-STEM). 4D-STEM reveals that 1T′ crystals grow from an originally nucleated kinetically favoured 1T phase which then transform into the thermodynamically stable 1T’ phase. Magneto-optic Kerr imaging reveals that the 1T’ $CrS_2$ crystals have soft ferromagnetic nature at low temperature. Our MOCVD growth of complex phases of 2D $CrS_2$ with long-range magnetic order paves the way for the scalable synthesis of 2D magnets for ultrathin magnetic memories for logic-in-memory applications and spintronics.

## Main

Two-dimensional materials that exhibit magnetic order[1, 2] have attracted considerable attention for the miniaturization of magnetic memories and spintronic devices in general. Chromium-based 2D magnetic materials were the first materials found to preserve their magnetism at the single layer limit ($CrI_3$, $Cr_2Ge_2Te_6$)[2, 3]. However, they are characterised by very low Curie temperatures (45-66 K), limiting their applications in cryogenic conditions. Layered Cr-dichalcogenides (TMDs) have been experimentally observed to exhibit robust magnetic behaviour close to room temperature (240-310 K) [4-7]. These materials exhibit structural polymorphism where the metal atoms within the individual atomic layers can have octahedral coordination (as found in the ‘T’ phases), trigonal prismatic coordination (as found the the ‘2H’ phase) [8-12] or more distorted structural coordination (as found in the T′, T″ and T‴ [13, 14] . Amongst these materials, $CrS_2$ has been predicted to display a diverse range of phase dependent properties, with the 1T, 1T′ and 2H phases predicted to be an antiferromagnetic metal, ferromagnetic semiconductor and non-magnetic semiconductor, respectively[15]. The 1T′ phase is of particular interest due to theoretically strain-dependent high Curie temperature, half-metallicity, and potential for harnessing the variable Curie temperature for applications in strain-controlled spin-valve logic gates[15, 16]. Moreover, the intercalation of additional Cr layers into $CrS_2$, yields the closely related $Cr_2S_3$ phase, which is found to be a non-layered ferrimagnetic semiconductor, with a Neel temperature of around 120 K. Such diverse electronic and magnetic properties make $CrS_2$ and other stochiometric Cr-S compounds promising candidates for investigating robust and persistent ferromagnetic order at and above room temperature in van der Waals nanosheet materials. The core challenge preventing exploitation of these promising magnetic properties is to selectively synthesise the different phases of $CrS_2$.

This is due to the similar thermodynamic stabilities of the different phases (0.34-0.76 eV per unit cell)[15], as well as competition from other stable stoichiometries such as $Cr_2S_3$[17-20], $Cr_3S_4$[21] and $Cr_5S_6$[20-22], with early attempts to synthesise $CrS_2$ having generated mixed phase samples with all 3 polymorphs (2H, 1T and 1T') experimentally observed [23]. Therefore, gaining full control, selectivity and understanding over the synthesis and properties of these materials remains elusive [23].

Here we demonstrate MOCVD synthesis of the previously unreported distorted octahedral 1T'crystal phase of $CrS_2$, as well as the selective and programmable synthesis of the 1T $CrS_2$ phase and $Cr_2S_3$. We reveal that the hydrogen concentration of the synthesis atmosphere together with the growth temperature, plays a key role in controlling the crystal phase, suggesting an interplay between kinetics and thermodynamics during synthesis. The phase structures were identified via polarized Raman spectroscopy and scanning transmission electron microscope (STEM) imaging and electron diffraction, correlated with density functional theory. The elemental stoichiometry of individual crystals was confirmed by STEM energy dispersive X-ray spectroscopy, while scanning electron diffraction (4D-STEM) has revealed that the synthesised 1T' crystals consist of differently oriented 1T′ nano domains within a hexagonal parent matrix. Magneto-optic Kerr effect (MOKE) microscopy reveals that the 1T' phase is magnetised at low temperature providing evidence of possible long-range magnetic ordering in the system. In addition, the intrinsic ferrimagnetism in $Cr_2S_3$ crystals is confirmed via M vs T measurements using a superconducting quantum interference device (SQUID). Our phase-selective MOCVD synthesis provides a scalable method for synthesizing 2D materials systems with a complex and sensitive phase behaviour, such as Cr–S compounds - paving the way towards creating a broad palette of magnetically-ordered 2D materials with appealing properties for integration in future memory, logic and memcomputing devices.

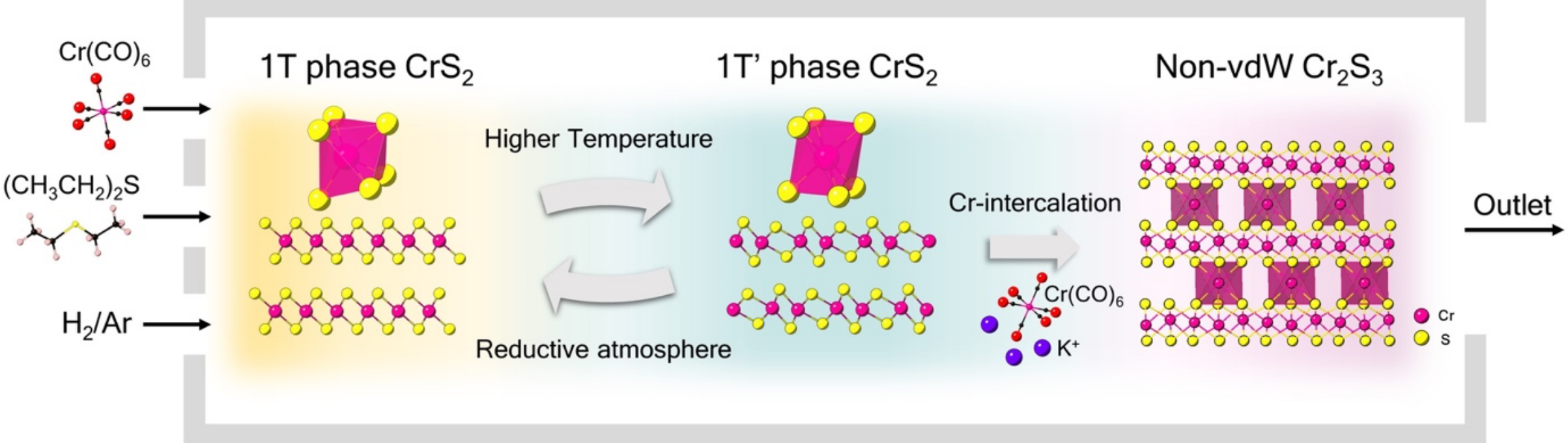

***Figure 1 Schematic diagram of the phase-selective synthesis of $CrS_2$ and $Cr_2S_3$ by MOCVD.*** *The direct syntheses of distinct 1T’ and 1T phases of layered $CrS_2$ and non-layered $Cr_2S_3$ are achieved by manipulating the growth conditions of MOCVD. A reductive atmosphere promotes the growth of 1T phase, while both thermal and kinetic factors drive the crystal phase transition between the 1T and 1T’ phases of $CrS_2$. Self-intercalation of Cr induced under a high Cr flux condition that catalyzed by specific alkaline metal ion ($K^+$) will generate the non-layered $Cr_2S_3$ phase. Pink and yellow spheres represent Cr and S atoms, respectively.*

**Phase-tuneable synthesis of layered $CrS_2$ crystals**

The direct synthesis of layered $CrS_2$ is realized by tuning the synthesis parameters during MOCVD (**Figure 1**). During synthesis, vaporized $Cr(CO)_6$ and $(CH_2CH_3)_2S$ precursors were flowed into a CVD growth chamber using a $H_2/Ar$ mixture carrier gas, with control of the flowrate and pressure. Different crystals shapes were obtained under distinct growth conditions (see Methods and **Supplementary Figures S1 and S9**). On sapphire substrates at 900 °C under a 5% $H_2/Ar$ atmosphere, optical micrographs reveal facetted, often triangular crystals were grown with a thickness of around 20-25 nm (**Figure 2a**) (Supplementary information). Decreasing the $H_2$ concentration to a 3% $H_2/Ar$ atmosphere at the same temperature, elongated crystal morphologies consisting of connected triangular shapes all having a similar orientation were observed (**Figure 2b**).

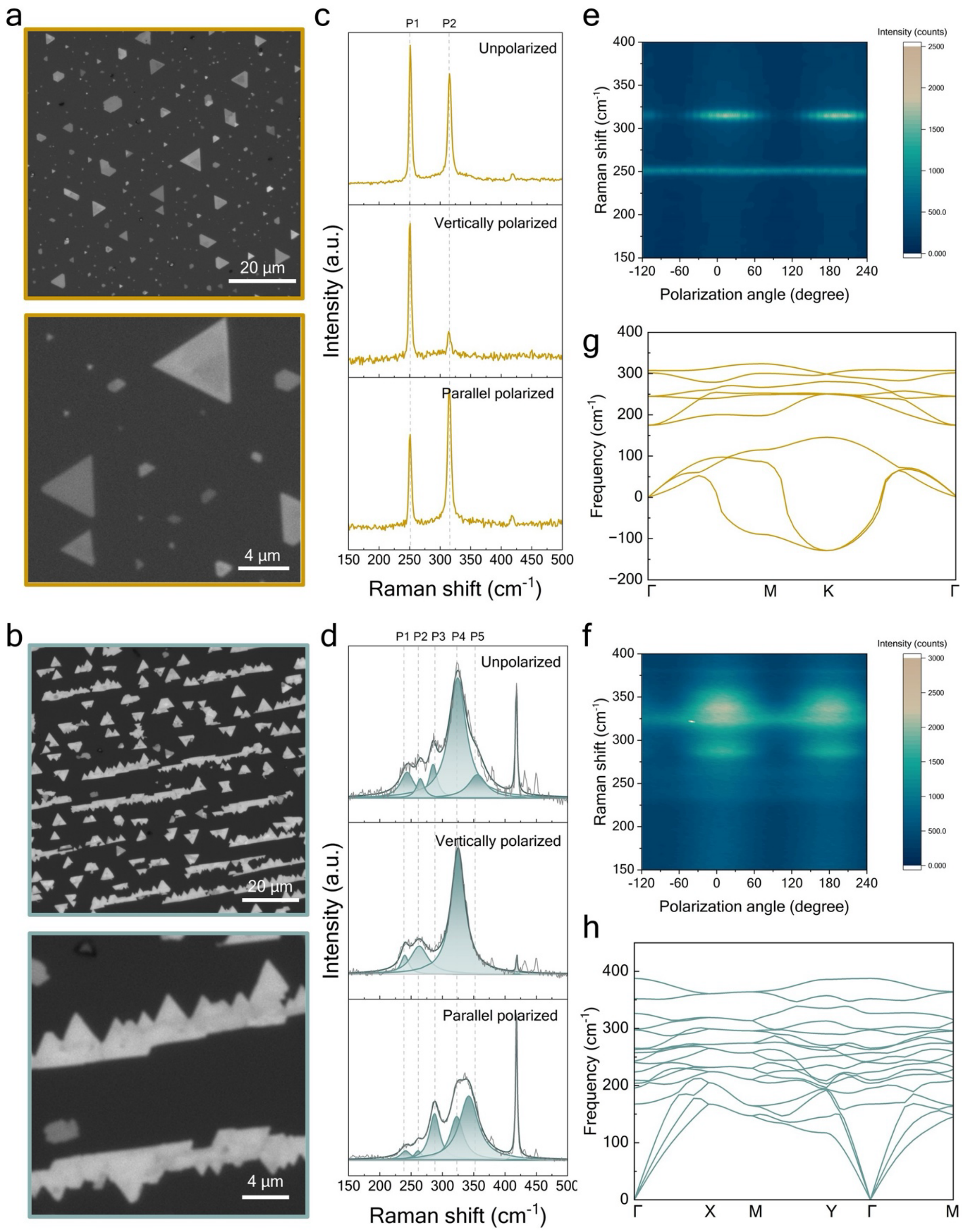

***Figure 2 Phase-selective MOCVD synthesis of $CrS_2$ polymorphs.*** *a, b Optical microscope images of the 1T phase (a) and 1T' phase (b) of $CrS_2$ synthesized on sapphire substrates at 900 °C under a 5% $H_2$/Ar and a 3% $H_2$/Ar atmosphere, respectively. c, d From top to bottom: Raman spectra for the 1T phase $CrS_2$ (c) and 1T' phase $CrS_2$ (d) in standard optics, and under*

*vertical and parallel polarized illumination configurations. Two sharp peaks in c are denoted as 1T-P1 and 1T-P2. The experimental spectra (solid grey line) in c are fitted with five Lorentzian peaks, denoted as 1T'-P1' to 1T'-P5' respectively. The Raman peaks at 378 $cm^{-1}$ and 418 $cm^{-1}$ belong to the sapphire substrate. e, f The angular dependence of the APR spectra of 1T $CrS_2$ (e) and 1T' $CrS_2$ (f) in the range of 150-400 $cm^{-1}$. g, h Calculated phonon dispersion spectra of $CrS_2$ crystals with the 1T phase (g) and 1T' phase (h).*

The two types of $CrS_2$ crystal morphology were investigated with Raman spectroscopy. The isolated triangular crystals (**Figure 2c**) grown under 5% of $H_2$ atmosphere, displayed two Raman modes at around 251.7 $cm^{-1}$ and 316.2 $cm^{-1}$ respectively, while the elongated crystals (**Figure 2d**) grown under 3% of $H_2$ atmosphere, had five Raman modes at ~241.9 $cm^{-1}$, ~259.6 $cm^{-1}$, ~291.6 $cm^{-1}$, ~331.8 $cm^{-1}$ and ~356.4 $cm^{-1}$ with the large difference pointing to the isolated triangular crystals having higher symmetry. To investigate the symmetry of each Raman mode we applied polarized Raman spectroscopy [24], comparing unpolarized and cross/parallel polarized conditions for both crystal morphologies (**Figure 2c,d**). It can be noticed that all the Raman modes are located between 200 $cm^{-1}$ and 400 $cm^{-1}$ similarly to the Raman spectra of the other group-VI TMDs [25] with the peak at ~ 418 $cm^{-1}$ assigned to an out-of-plane vibrational peak of the sapphire substrate [26]. For the isolated $CrS_2$ particles the two peaks present in parallel polarisation are referred to as $P_1$ and $P_2$, while in cross polarization the intensity of $P_2$ is negligible (**Figure 2e)**. The Raman spectra of the crystals with the elongated morphology can be fitted with five Lorentzian peaks, which are referred to as $P_1$'-$P_5$', respectively (**Figure 2c**). The positions of $P_1$', $P_2$' and $P_4$' peaks have nearly the same intensity for cross and parallel polarized light while $P_3$' and $P_5$' peaks disappear under cross polarization suggesting they are associated with a characteristic out-of-plane *A* mode, as the *A* modes are forbidden in cross polarization [24]. Since small changes in the intensity of the peaks can cause uncertainties in distinguishing the phonon modes under a single cross or parallel polarization, angle-resolved polarized Raman (APR) was performed, varying the angle of polarization of the incident light, $\theta$, from 0 to 360º.

APR plots of the Raman intensity versus the peak position at different polarisation angles for the isolated triangle $CrS_2$ morphology (Figure 2a) and the elongated $CrS_2$ morphology (Figure 2b), are shown in **Figures 2e and f** respectively (equivalent contour graph are given in Supplementary Information Figures S5 and S6). The normalized intensities were fitted for each peak separately according to the Raman tensor of $CrS_2$ (calculation details are given in **Supplementary Information Note 1**). We consider first the APR plots for the isolated $CrS_2$ morphology, where the peak intensity of $P_1$ remains constant in the whole-angle polarization measurement while the peak intensity of $P_2$ shows an obvious angular dependence, consistent with the polarised and cross polarised measurements. The polar plot of $P_1$ can be fitted with an almost perfect circle (Supplementary Information Figure S6), suggesting this peak should be described as an in-plane vibrational *E* mode resonance. On the other hand, $P_2$ has the profile of a $\cos^2\theta$ function, suggesting this peak has the *A* symmetry phonon mode nature. We next consider the APR of the elongated $CrS_2$ morphology (**Figure 2f**). According to the shape of the polarization dependent intensities, the polar plots of $P_3$' and $P_5$' have the shape of a $\cos^2\theta$ function, which is a distinct feature of the *A* symmetry. While the polar plots of $P_1$' and $P_2$' show a circular profile, corresponding to the expected behaviour of the in-plane *E*-like phonon modes. $P_4$' shows a convolution of a circle and a $\sin^2\theta$ function, although the intensity modulations are weak, likely due to overlap of different intensities.

To understand the two distinct crystal structures of the $CrS_2$ crystals that give rise to these different Raman signatures, density functional perturbation theory (DFPT) calculations were preformed to determine the phonon dispersion for the three most likely crystal structures of $CrS_2$ (the trigonal prismatic 2H phase, the octahedral 1T phase and the distorted octahedral 1T' phase). The computed phonon frequencies of the 1T phase $CrS_2$ (at 244.9 and 307.4 $cm^{-1}$, see **Figure 2g**) are a good match to the experimental Raman modes for the isolated triangular morphology of $CrS_2$ grown in 5% $H_2$/Ar condition (peaks at 251.2 and 315.7 $cm^{-1}$). However, the free-standing 1T phase of monolayer $CrS_2$ is calculated to be dynamically unstable as indicated by the negative phonon frequencies at the high-symmetry points. The computed Raman-active optical phonon frequencies at the Γ point for the monolayer 1T′ phase of monolayer $CrS_2$ (at 214.5, 221.1, 239.6, 268.7, 296.9, 322.6, and 357 $cm^{-1}$ in **Figure 2h**) are in good agreement with the experimental peaks observed in the Raman spectra for the elongated $CrS_2$ morphology synthesised at lower $H_2$ concentration. Note that small differences in the absolute phonon frequencies are expected as the structural relaxations are performed with DFPT. For completeness the computed Raman-active optical phonon frequencies for the 2H phase of $CrS_2$ is also shown in Supplementary Information Figure S7. These show a good agreement with the Raman modes of 2H $CrS_2$ reported in the literature at 391.4 and 405.6 $cm^{-1}$ [23] but do not correspond to our experimental data. A fit to the $Cr_2S_3$ structure can also be ruled out based on the literature reports for the Raman modes of this material (at ~ 176.4, 252.4, 285.1, and 364.0 $cm^{-1}$ [27, 28]. To fully understand the tuneability between these two phases, we have performed a range of syntheses experiments varying the temperature and the hydrogen content to build a phase diagram of syntheses conditions leading to the formation of either the 1T' or the 1T phase (Figure 3a). Optical images and Raman spectra were used to assign the phase of the synthesised crystals (SI Figure S9). These results show that the 1T' phase of $CrS_2$ is obtained at temperatures between 900 °C up to 950 °C, in an atmosphere of pure Ar or low $H_2$ concentration (3%) (Condition 1). While the growth of the 1T phase $CrS_2$ is more favourable under a higher $H_2$ concentration at lower temperatures in the range of 800-850 °C and (Condition 2). Such findings suggest that the 1T phase tends to form in synthesis condition which favour kinetic stabilization (lower T and with hydrogen stabilization), which is consistent with our prediction of the dynamical instability of the monolayer 1T phase of $CrS_2$ demonstrated by the negative phonons present in our DFPT calculations. While the stable 1T' phase is suggested to form in conditions where kinetic barriers are easier to overcome enabling thermodynamic stabilisation (higher temperature). We observe there is an interval region around 850-900 °C for low $H_2$ concentration (<3%) where both crystal phases are found to be present (**Figure 3a**). The Raman spectrum in **Figure 3b** shows the Raman spectrum from $CrS_2$ synthesised in this regime, which contains characteristic resonance modes from both 1T' and 1T phases, suggesting the existence of both crystal phases in the sample. X-ray diffraction (XRD) was also used to confirm the phases synthesised for different conditions in the phase diagram with the experimental XRD patterns of the pure phase 1T' and 1T $CrS_2$ materials in **Figure 3c** found to be consistent with simulated XRD patterns (see Figure S8). Specifically, the (002) peak of the as-synthesized 1T' phase $CrS_2$ located at around 2 theta=12.81° can be clearly distinguished from the (001) peak of as-synthesized 1T phase at 2 theta=13.70°. The mixed phase region shows XRD peaks from both 1T' and 1T $CrS_2$ phases with the relative intensities of the characteristic peaks for each phase able to shed light on the relative abundance of the two phases in the material. Such relative quantification reveals a gradual transition from the 1T phase to the 1T' phase with increasing temperature across the mixed phase region, enabling the synthesis of phase mixtures of $CrS_2$ if required. Hence, a selective crystal phase transition of $CrS_2$ can be feasibly achieved by direct controlling synthesis temperature and atmosphere during MOCVD.

To fully understands the tuneability between these two phases, we have performed a range of syntheses experiments varying the temperature and the hydrogen content to build a phase diagram of syntheses conditions leading to the formation of either the 1T' or the 1T phase (Figure 3a). Optical images and Raman spectra were used to assign the phase of the synthesised crystals (SI Figure S9). These results show that the 1T' phase of $CrS_2$ is obtained at temperatures between 900 °C up to 950 °C, in an atmosphere of pure Ar or low $H_2$ concentration (3%) (Condition 1). While the growth of the 1T phase of $CrS_2$ is more favourable at low temperatures within the range of 800-850 °C and under a higher $H_2$ concentration compared to the 1T' phase (Condition 2). Such findings suggest that the 1T phase tends to form in synthesis condition which favour kinetic stabilization (lower T and with hydrogen stabilization), which is consistent with our prediction of the dynamical instability of the monolayer 1T phase of $CrS_2$ demonstrated by the negative phonons present in our DFPT calculations. While the stable 1T' phase does form in conditions which establish a thermodynamic equilibrium (high temperature). We observe there is an interval region around 850-900 °C for low $H_2$ concentration (<3%) where both crystal phases are found to be present (**Figure 3a**). The Raman spectrum in Figure 3b includes characteristic modes of both 1T' and 1T phases, which reveals the existence of both crystal phases. The presence of both phases is also confirmed by X-ray diffraction (XRD) (Figure 3c). The experimental XRD patterns in **Figure 3c** are consistent with the simulated XRD patterns of pure 1T' and 1T phases of $CrS_2$ crystals (FigureS11). The (002) peak of the as-synthesized 1T' phase $CrS_2$ located at around 2 theta=12.81° can be clearly distinguished from the (001) peak of as-synthesized 1T phase at 2 theta=13.70°. The mixed phase region shows XRD peaks from both 1T' and 1T $CrS_2$ phases with the relative intensities of the characteristic peaks for each phase able to shed light on the relative abundance of the two phases in the material. Such relative quantification reveals a gradual transition from the 1T phase to the 1T' phase with increasing temperature across the mixed phase region, enabling the synthesis of phase mixtures of $CrS_2$ if required. Hence, a selective crystal phase transition of $CrS_2$ can be feasibly achieved by direct controlling synthesis temperature and atmosphere during MOCVD.

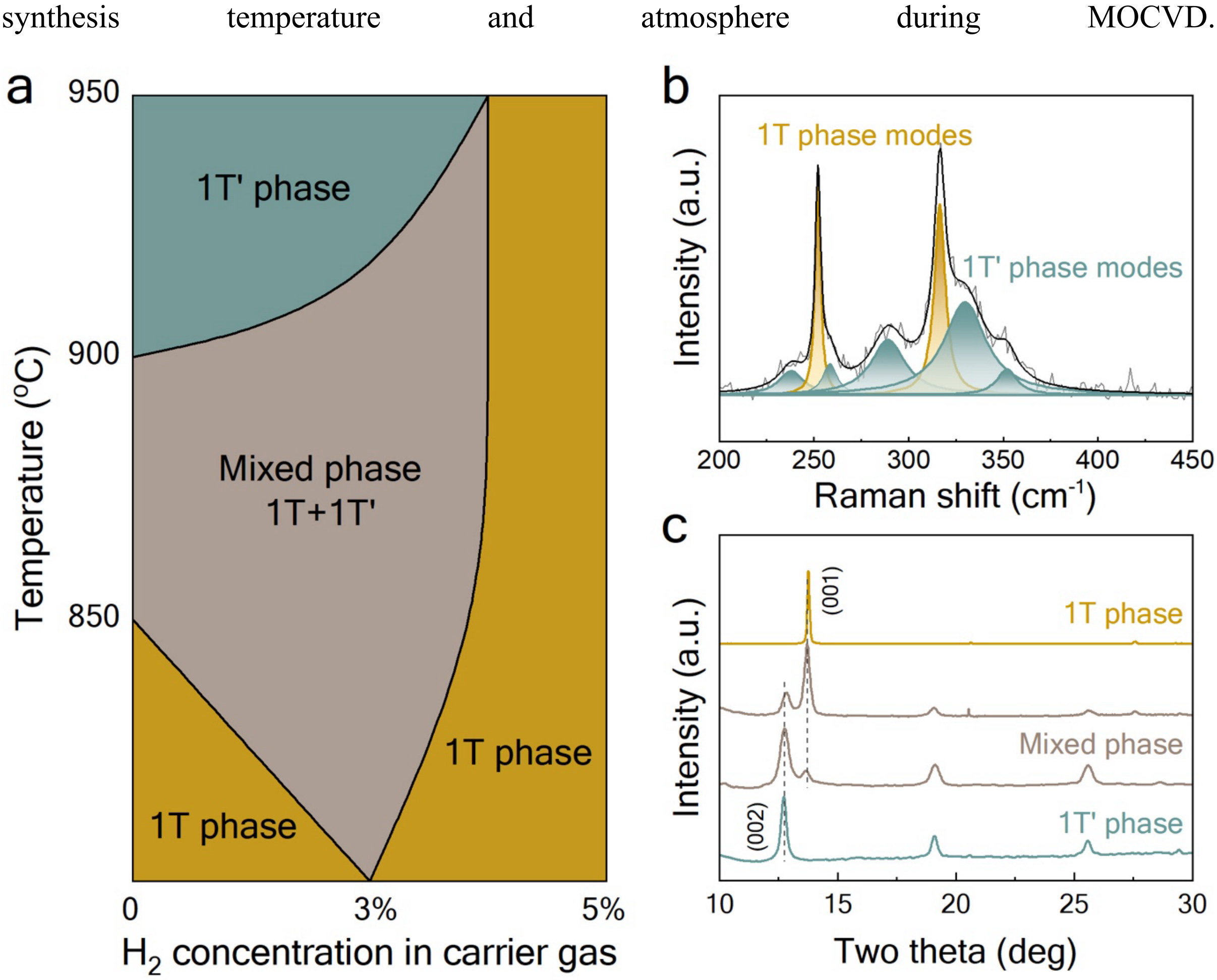

***Figure 3 Phase transitions between 1T' phase and 1T phase $CrS_2$.*** *a, Phase diagram for the synthesis of $CrS_2$ as a function of growth temperature and $H_2$ concentration in argon carrier gas. b, Raman spectrum of a mixed 1T'+1T phase $CrS_2$ sample. c, XRD patterns of 1T' phase, 1T phase and mixed phase of $CrS_2$.*

Scanning electron microscopy (SEM) and scanning transmission electron microscopy (STEM) with energy dispersive X-ray spectroscopy (EDS) was also used to confirm the successful synthesis. STEM EDS of 1T $CrS_2$ when grown using condition 2, revealed a Cr:S ratio of 0.49 with a standard deviation of 0.1 in agreement with the stoichiometrically expected value of 0.5 (**Figure 4a**, and SI Figures S11) while the 1T' flakes grown using condition 1 had a slightly higher Cr:S ratio of $0.59 \pm 0.11$. To gain more insights into the structure and uniformity of the synthesised 1T and 1T' $CrS_2$ phases, their atomic structures were further evaluated using atomic resolution high angle annular dark field (HAADF) STEM imaging and selected area electron diffraction (SAED). The morphologies of flakes grown under both conditions are as expected from the optical and SEM images with facetted edges and frequent 60° or 120° apices leading to triangular or hexagonal flakes, as is commonly observed in CVD growth of other TMDs [29-31]. The atomic imaging of the crystals grown using condition 2 show clear hexagonal symmetry, with an atomic column intensity distribution expected for the 1T phase (**Figure 4b**, also see STEM simulations of different phases in SI Fig S12a and S14). SAED patterns further reveal the isolated crystals are single crystals of 1T phase $CrS_2$ (SI Fig S12b-d). In contrast, the atomic structure of the flake grown using condition 1 shows a distinctive breaking of hexagonal symmetry associated with the 1T' distortion (**Figure 4e**), which is also visible as the additional lattice spots in the correspond ing fast Fourier transform (FFT) of the image (inset in **Figure 4e**). The presence of the 1T′ distortion in the $[11\underline{2}0]$ direction is additionally evident in the corresponding SAED pattern, where the characteristic additional diffraction spots at half intervals arise due to the additional periodicity in the unit cell created by the dimerisation of adjacent Cr ions (**Figure 4f**). Interestingly, although some areas of the crystals show the expected SAED pattern shown in **Figure 4f**, in other areas multiple orientations of the 1T′ phase can be observed in a single SAED (SI Fig S15). While the SAED retains the underlying hexagonal symmetry for the whole flake, the direction of the 1T′ distortion varies between three available $< 11\underline{2}0 >$ directions in the hexagonal crystal system. Measurement of the diffraction patterns give lattice parameters of a= b=3.38 Å, γ=120° for the hexagonal unit cell of the 1T phase, while values of a=3.42 Å and b=5.87Å, γ=90° are determined for the tetragonal unit cell of the 1T' phase. This agrees with the established trend observed in other 1T' phases of TMDs where $b \approx \sqrt{3}a$, when considering a rhombohedral basis rather than a hexagonal unit cell.[32]

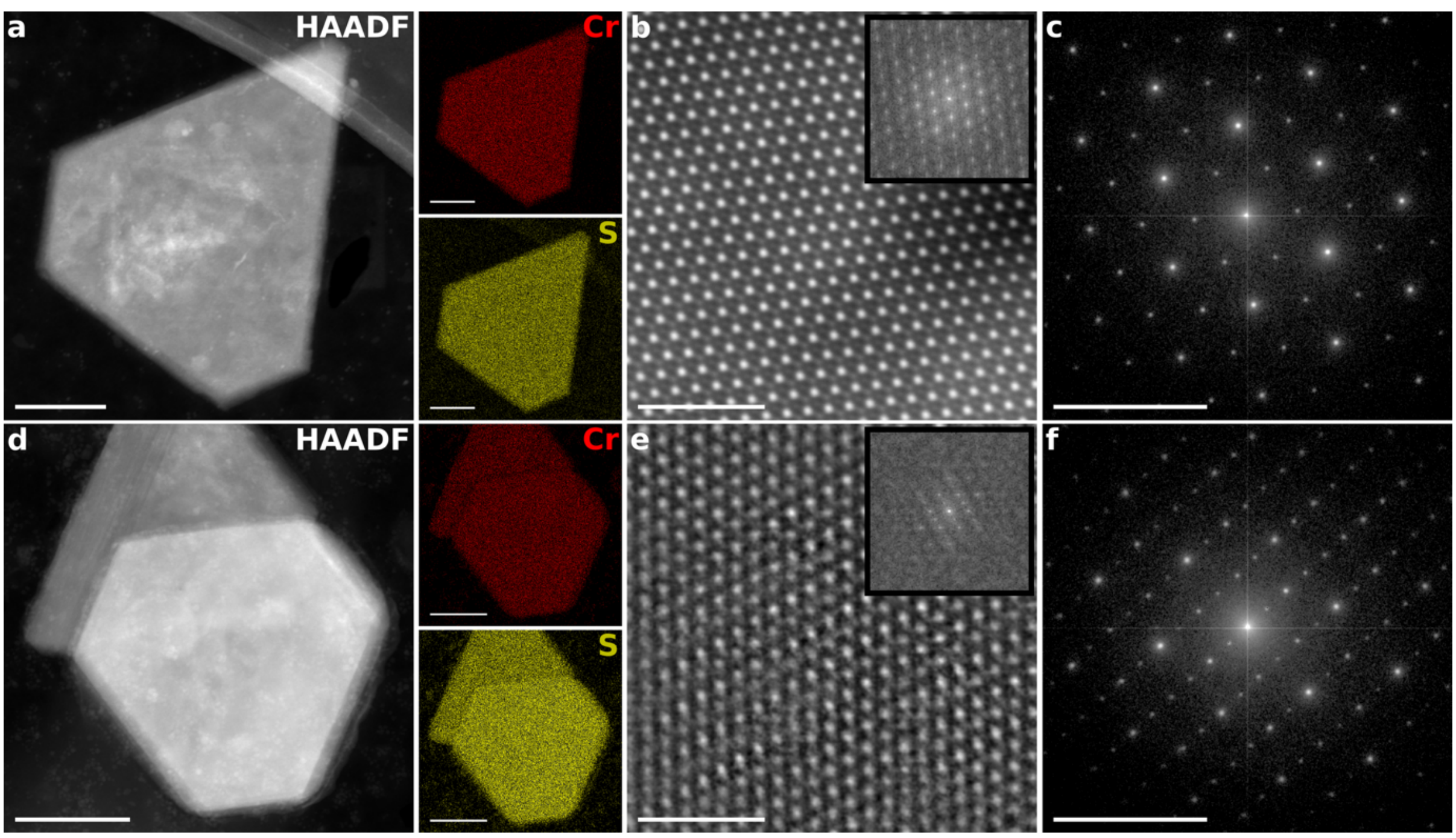

***Figure 4 TEM characterization of crystal structures of $CrS_2$***. *a, d HAADF STEM images and STEM EDS maps of Cr and S in a 1T phase crystal (a) and 1T' phase crystals (d) $CrS_2$. b, e HAADF STEM images of atomic structures of 1T $CrS_2$ (b) and 1T' $CrS_2$ (e). Insets are FFTs of the corresponding atomic resolution images, with the 1T' showing the additional periodicity induced by the distortion. c, f SAED patterns of a 1T flake (e) and a 1T' flake (f). Scalebars in a and d are 100 nm, in b and e are 2 nm and in c and f are $10nm^{-1}$.*

To better understand the variability of the 1T' distorted periodicity seen in the SAED patterns which have a ~200 nm diameter sampling region, scanning precession electron diffraction (SPED) was used to map the nanoscale distribution of different orientations of the 1T′ in flakes grown using condition 1. Acquiring a 4D-STEM data set using a precessed STEM probe with a diameter of ~2nm on the specimen, reveals the existence of a complex network of nanodomains that could be assigned to the 1T' phase with three different orientations (**Figure 5b,**) or to regions where two different domain orientations overlap (**Figure 5c**). Further details of the domain assignment can be found in methods and SI Figure S18. Each orientation has the 1T' distortion along one of the three $< 11\underline{2}0 >$ directions of the hexagonal lattice as illustrated in the corresponding diffraction patterns and unit cell schematics (**Figure 5h-j** and **l-m** respectively). The underlying hexagonal diffraction spots are invariant across the flake, consistent with the SAED, indicating an underlying parent matrix with hexagonal symmetry for all the domains. Where two different 1T' domain orientations meet the diffraction pattern has unusual streaking present between the parent hexagonal lattice spots orthogonal to the distortion direction (**Figure 5g,k, SI Figure S16**), which is likely to be due to the presence of stacking faults where the different orientation domains overlap in the [0001] direction. Diffuse streaking of this kind is known to be associated with crystallographic disorder likely due to overlapping of different orientations of 1T′ in the [0001] direction of the crystal as illustrated in SI Figure S17. It was further observed that phase orientations with the distortion aligned perpendicular to one of the free edges were more commonly observed in proximity to that edge, suggesting that the edge proximity provides a route to reduce strain from the 1T' lattice distortion (see SI Figure S18 for illustration of mapping 1T' orientation preference as a function of distance from edge).

These observations provide some interesting insights into the growth mechanism of the 1T’ crystals, suggesting that the crystal originally nucleates in the 1T phase, which dictates the underlying hexagonal symmetry, before transforming to 1T′. The phase transformation induces strain in the lattice, which will be reduced be the presence of a nearby edge perpendicular to the direction of the induced lattice strain. We therefore hypothesise that the edge regions of the 1T $CrS_2$ crystal may provide lower energy sites for the 1T’ to nucleate. If many 1T’ domains nucleate independently, these will then have to compete for the available 1T crystal during growth, which could result in the complex nanoscale domain structure we observe experimentally.

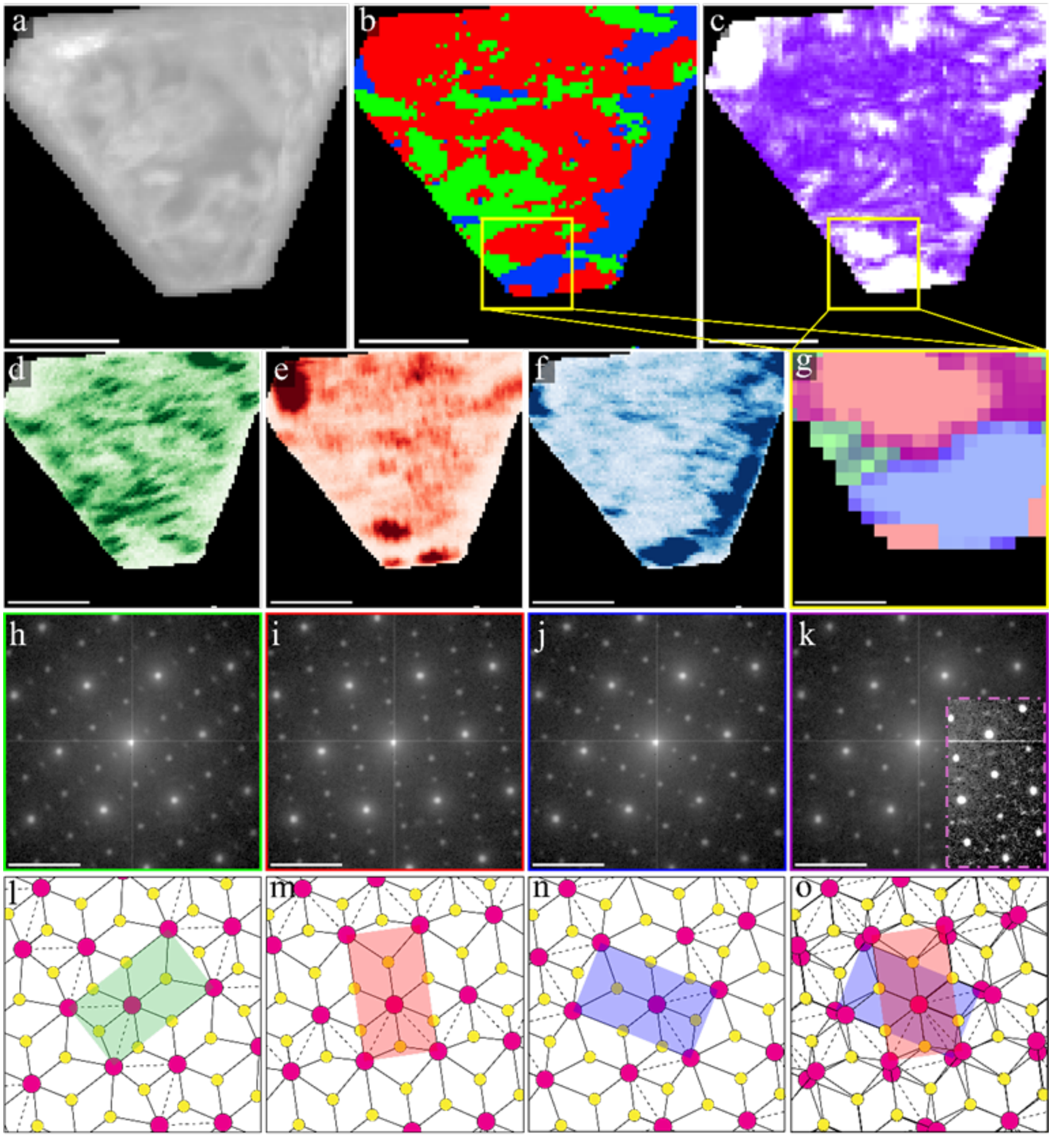

***Figure 5 1T' crystal orientation mapping using 4D-STEM.*** *a, Virtual bright field image of 1T' $CrS_2$ flake grown using condition 1. b, Orientation map of 1T' crystals characterised by atomic distortion along different* $[11\bar{2}0]$ *directions of the hexagonal parent lattice. c, Map of the lines of low-confidence orientation assignment corresponding to diffuse diffraction pattern assigned to stacking faults between orientation domains. d-f, Maps of individual 1T' phase orientations across the flake, with corresponding diffraction patterns shown below in h-j. g, Overlayed enlarged regions from b and c, showing presence of all three orientations at a corner, with each domain surrounded by mixed phase stacking faults (purple), the corresponding diffraction pattern of which is shown in k. Inset to k has been contrast adjusted to enhance visibility of weak scattering features. l-o, Atomic schematics of different orientations of 1T', with boxes showing the tetragonal unit cells of the 1T' phase with different orientations (l-n), and proposed structure resulting in a stacking fault formed by differently oriented 1T' crystals overlapping along the* $[0001]$ *direction of the crystal (o). Scale bars in a-f, g, and h-k are 200 nm, 50 nm and 5* $nm^{-1}$*, respectively. 1T' distortion in l-o is exaggerated and not to scale*

**Synthesis of self-intercalated $Cr_2S_3$ crystals**

We now consider the ability to synthesise $Cr_2S_3$ using MOCVD since to the best of our knowledge, direct synthesis of $Cr_2S_3$ has been only realized through conventional CVD using powder precursors.[28] Here, we engineered the synthesis of $Cr_2S_3$ through a facile self-intercalation process of $CrS_2$ during using MOCVD and observed a phase transition from 1T' to $CrS_2$-mixed 1T' $CrS_2$ and $Cr_2S_3$ to pure $Cr_2S_3$ crystals. As shown in the schematic diagram in **Figure 6a,b** this phase transition to $Cr_2S_3$ is achieved by introducing KCl [45] during the synthesis, whereas simply increasing the Cr:S ratio in the presence of NaCl did not result in synthesis of $Cr_2S_3$. We hypothesize that the role of the potassium ion, being larger than sodium is to lower the energy barrier for the intercalation of Cr between the layers of $CrS_2$ [45]. Polarized and unpolarised Raman spectra for the crystal synthesised in the presence of KCl are shown in Figure 6c-e and provide evidence of a crystal structure distinct from the 1T' or 1T phase $CrS_2$. Six well-pronounced Raman modes are observed in the unpolarised condition and are referred to as $P_1 - P_6$. Based on the dependence of the Raman tensor on polarization, the active Raman modes at 256.8 $cm^{-1}$ ($P_1$), 303.6 $cm^{-1}$ ($P_3$) 338.5 $cm^{-1}$ ($P_5$) (**Figure 6d**), are assigned to be in-plane E modes while the modes at 281.6 $cm^{-1}$ ($P_2$), 316.1 $cm^{-1}$ ($P_4$) 364.8 $cm^{-1}$ ($P_6$) are out-of-plane A modes.

A similar assignment of the A and E modes of $Cr_2S_3$ has been reported in the literature [46,47] . Further confirmation of the successful synthesis of the $Cr_2S_3$ structure has been gained via XRD and STEM imaging (SI Figures S18 and S19).

Although the presence of KCl was found to be a prerequisite for synthesis of $Cr_2S_3$, this phase is not always obtained when KCl is in the system. For example, when synthesizing for 60 min under S-rich conditions (flux ratio Cr:S=1:2), a pure 1T' $CrS_2$ phase is obtained even with the presence of KCl. At a longer synthesis time of 120 min under the same conditions, the 1T' $CrS_2$ phase is mainly retained although the flake size increases from around 1 μm to around 3-5 μm and $Cr_2S_3$ nuclei start to form around the $CrS_2$ flakes (see SI Figure S21). Under a more Cr-rich environment (flux ratio Cr:S=1:1), we find that a mixture of 1T' $CrS_2$ and $Cr_2S_3$ phase are synthesised, where 1T' $CrS_2$ flakes are only partially self-intercalated to create layers of $Cr_2S_3$ phase in the 1T' $CrS_2$ phase host (SI figure S19). At longer synthesis times of 120 minutes, crystals of $C_2S_3$ grow around triangular 1T' $CrS_2$ flakes (Figure S21, similar to the case in low Cr-flux condition). A pure $Cr_2S_3$ phase requires a longer 180 min synthesis under high Cr-flux (flux ratio Cr:S=1:1) with KCl promotion where the as-obtained pure $Cr_2S_3$ phase shows a good homogeneity under optical microscopy and atomic force microscopy (AFM) (SI Figure S22).

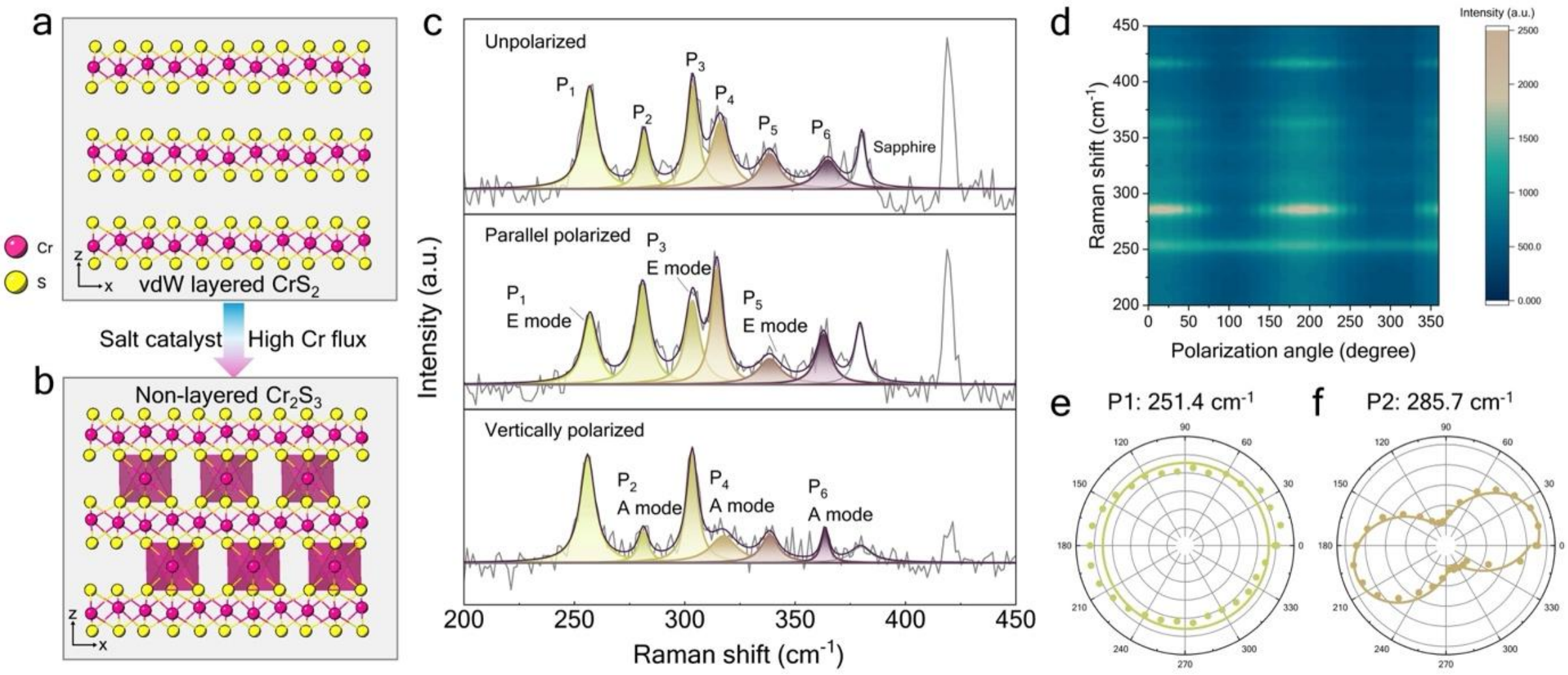

***Figure 6 Synthesis of non-layered $Cr_2S_3$ from 1T' phase $CrS_2$ by self-intercalation of Cr.*** *a, b Schematic of the conversion from vdW layered $CrS_2$ to non-layered $Cr_2S_3$ though self-intercalation of native Cr atoms catalysed by KCl salt. c, From top to bottom: polarized Raman spectra of in standard optics, vertically and parallel polarized configurations. The experimental spectra (solid grey line) are fitted with six Lorentzian peaks, denoted as P1 to P6*

*respectively. d, The angular dependence of APR spectra of $Cr_2S_3$ within the range of 150-400 $cm^{-1}$. e, f Polar plots of normalized intensities of the two Raman modes (P1 and P2) of $Cr_2S_3$ as a function of incident polarization angle. In the polar plots, the spheres are experimental data and solid lines are fittings.*

**Magnetic properties**

Here we examine the magnetic properties of both Cr2S3 and CrS2. The magnetisation as a function of temperature is measured for $Cr_2S_3$ crystals is measured by a superconducting quantum interference device (SQUID). We perform zero-field-cooling (ZFC) and field-cooling (FC) measurements to discern the magnetic ordering temperature. We perform ZFC-FC measurements for three different applied magnetic fields for the FC portion 100 Oe, 1000 Oe, and 10k Oe (Figure 7) –.The observed maximum is sharpened in the 10000 Oe data. For the FC and ZFC data with 100 Oe, the moment is noisier because the moment is comparable to the noise floor of the magnetometer, but there is clear divergence of ZFC and FC with the onset of non-Curie-Weiss behaviour at around ~ 70K. The observations are consistent with low temperature ferrimagnetic order behaviour. The characteristic maximum of magnetization is field dependent, observed in the range 40 K – 70K, which is lower than what has been reported so far in the literature (~76 K) [19]. This different could be attributed to differences in anisotropy from different sample morphology and aspect ratio (e.g. finite size/shape anisotropy effects from the nanoscale crystallites), or to a possible small substochiometric ratio of Cr in our $Cr_2S_3$.

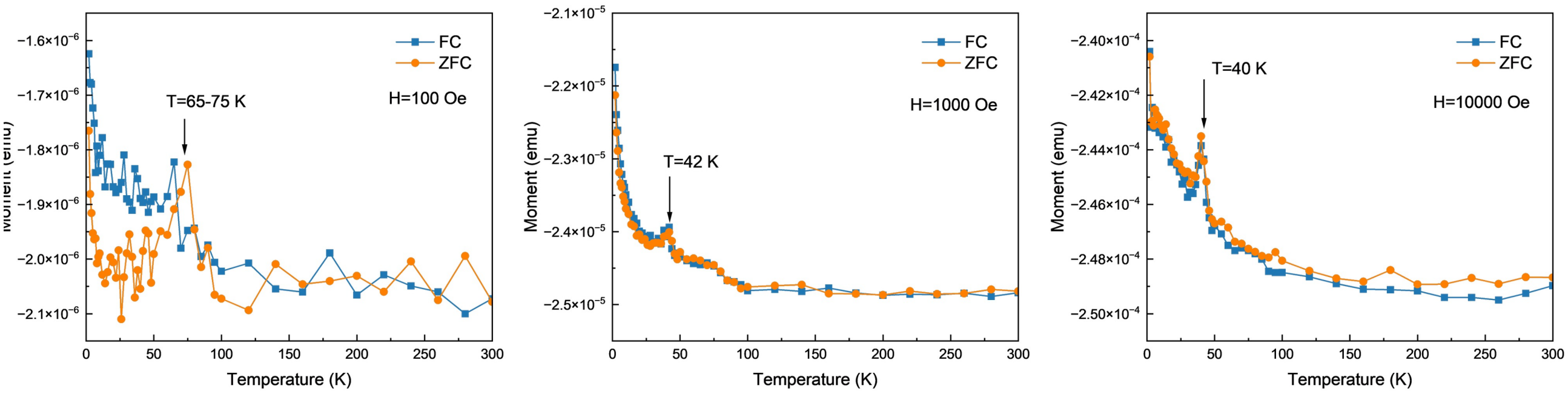

***Figure 7*** *– Temperature dependence of magnetization of $Cr_2S_3$ crystals on sapphire substrate. FC-ZFC curves of $Cr_2S_3$ crystal under an external magnetic field of 100 Oe (**a**) and 1000 Oe (**b**), and 10000 Oe (**c**), respectively.*

We now investigate the magnetic properties of $CrS_2$. Single crystal flakes of as-grown $CrS_2$ crystals (reflection images shown in Supplementary Figure S18) were measured optically by polar MOKE at 4.5 K (probing the out-of-plane component of magnetisation), with a clear magnetic signal observed which persists at zero field showing ferromagnetic order. At 4.5 K, the clear polar MOKE signal coming from the crystals stands out from the weaker paramagnetic signal of the sapphire substrate, and switches from positive to negative when reversing the external axial magnetic field (applied perpendicular to sample/substrate plane) from 3T to -3 T (panels (a) and (b) of Figure 8 respectively). Upon further removing the external field, the contrast diminishes (Figure 8 (c)), showing a weak yet sizeable remanent polar MOKE signal of a magnetised state. The remanent MOKE signal is then lost when the sample is warmed up to 285 K (Figure 8 (d), showing transition to a non-magnetic phase.

Therefore, we conclude that the 1T' $CrS_2$ crystals display a clear magnetic contrast at low temperatures, high out-of-plane external magnetic fields, and signs of a remanent magnetic state at 0 T that vanishes upon warming up the sample to room temperature.

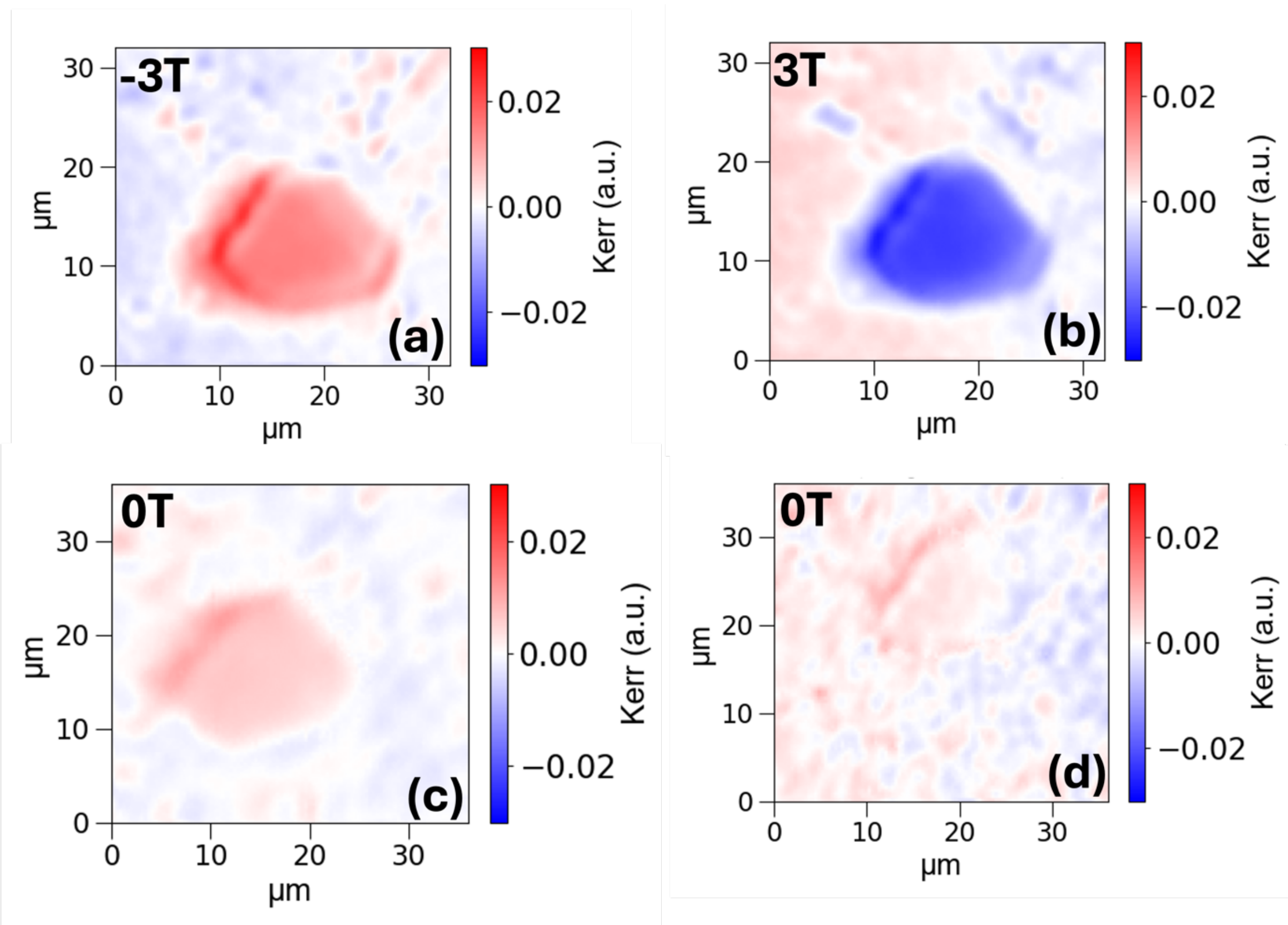

***Figure 8*** *Polar MOKE signal of a $CrS_2$ 1T' crystal grown on a sapphire substrate. Measurements were taken at base temperature (4.5 K) in an out-of-plane magnetic field of -3 T (a), 3 T (b), and 0 T (c). (d) MOKE micrograph of the same crystal at 285 K and 0 T.*

## Conclusions

We have demonstrated the MOCVD synthesis of the distorted octahedral (1T') phase of $CrS_2$ with a long-range magnetic order. The nanoscale analysis of this crystal phase revealed a novel disordered structure characterised by a complex network of nanodomains with different orientations of the 1T' phase of $CrS_2$ present within a host single crystal parent matrix. The characterisation suggests a growth mechanism comprised of initial nucleation and growth of 1T $CrS_2$ that is then transformed to the 1T′ phase. Moreover, we show that phase and stoichiometry are highly dependent on growth conditions and span a broad, compositionally variable phase diagram which include the 1T phase of $CrS_2$, the 1T' phase and a different stoichiometry of Cr-S, which is the ferrimagnetic $Cr_2S_3$. In this work, we have reported a new magnetic 2D material grown with a scalable technique opening new avenues in integrating 2D magnetic materials in memory devices for use in logic-in-memory and spintronic devices.

## Methods

**Synthesis of 1T' and 1T phases $CrS_2$.** The tunable synthesis of 1T' and 1T phases of $CrS_2$ is conducted via metal-organic chemical vapor deposition (MOCVD). Chromium hexacarbonyl (CHC, $Cr(CO)_6$) (STREM Chemicals UK, Ltd.) and diethyl sulphide (DES, $(CH_3)_2S$) (Sigma-Aldrich) which have equilibrium vapor pressures near room temperature are used as the organic precursors and sealed in two stainless steel bubblers (Strem Chemicals UK, Ltd.). The bubblers are in a 30 °C water bath (LAUDA Technology Limited) to ensure continuous vapor generation. Argon (Ar) with a certain amount of $H_2$ (3% and 5%) is used as the carrier gas to transport the precursor vapor to the reaction chamber. The gas flowrate and pressure are automatically monitored and controlled by the mass flow controllers (Aalborg Instruments) and pressure controllers (Bronkhorst High-Tech B.V.). The sapphire substrates were washed with acetone and deionized water before being placed in the center of the quartz tube. A crucible of sodium chloride (NaCl, Sigma-Aldrich) (and KCL instead for $Cr_2S_3$) is placed upstream as a desiccant to dehydrate the reaction chamber.

**Morphological characterizations.** The morphology of MOCVD-grown $CrS_2$ crystals is characterised with atomic force microscopy (AFM, Asylum MFP-3D, Oxford Instruments) and scanning electron microscopy (SEM, Zeiss Auriga). AFM measurement uses the AC air topography mode. The accelerating voltage used for the SEM imaging is 5 kV and the working distance is 5-10 mm. SEM-EDS mapping is performed at 10 kV using an Oxford Instruments x-act PentaFET Precision detector.

**Raman spectroscopy.** Raman spectroscopy and polarized Raman measurements are used to demonstrate the crystal phases of $CrS_2$ and assign the characteristic vibration modes. The Raman spectra are collected with a Renishaw inVia spectrometer, the excitation laser wavelength is 532 nm and the grating resolution is 2400 lines/mm. Polarized measurement is conducted by adjusting the polarized configuration of the incident light and scattered light.

**Chemical composition and crystal structures.** Chemical composition is determined by X-ray photoelectron spectroscopy (XPS, ThermoScientific K-Alpha$^+$) equipped with a monochromatic Al $K_\alpha$ X-ray source. The pass energies used for survey spectra and core levels are 200 eV and 20 eV respectively. X-ray diffraction (XRD, MPD PANalytical) equipped with a Cu $K_\alpha$ X-ray source and transmission electron microscopy (TEM, JEOL JEM-2100Plus) are used to demonstrate the crystal structures of $CrS_2$. For TEM measurement, the MOCVD grown $CrS_2$ is transferred from the sapphire substrates to commercial Cu grids covered with a holey carbon support film. Polystyrene (PS) is used as the support to peel off the $CrS_2$ from sapphire following a transfer process outlined in the literature[52].

**STEM Imaging and Analysis**: Atomic resolution high angle annular dark field (HAADF) STEM images were collected using an FEI Titan "ChemiSTEM" (G2 80–200) operated at 200 kV with an aberration-corrected probe, a 21 mrad probe convergence angle, and a HAADF inner angle of 64 mrad. STEM EDS spectroscopy data was collected using a FEI Talos F200A TEM at an operating voltage of 200 kV, equipped with a Super-X EDS detector (4× windowless silicon drift detector). SAED was collected with a 0.21 ± 0.1 μm SAED aperture with diffraction patterns for SAED and 4D-STEM collected using a Quantum Detectors Merlin Quad detector with 512 × 512 pixels. The cross visible in the diffraction data is a result of the missing pixels at the join between the Merlin detector's 4 chips (each 256 × 256). STEM simulations were performed using open-source python package abTEM[24] matching the experimental parameters of the FEI Titan 'ChemiSTEM'. Image analysis and filtering was performed using open-source python package Hyperspy version 2.3.0 [hyperspy v2.3.0 (Zenodo, 2025).

**Cross sectional sample preparation:** Cross-sectional lamellae were fabricated using an FEI Helios NanoLab 660 dual beam focused ion beam (FIB) SEM. Before FIB miling the samples were coated with 20 nm of carbon. The target slice for lift out was then coated with a 2 μm thick strip of Pt via electron beam then ion beam deposition, before a ion milling at 30 kV to create a slice cross section. The slice was then thinned at voltages down to 5 kV to achieve electron transparency.

**First-principles calculation.** The theoretical first-principles calculations are performed using density functional theory (DFT) with the PBE exchange-correlation functional-Norm-Conserving-Vanderbilt pseudopotentials phases (2H, 1T', 1T) of the monolayer $CrS_2$ and bulk $CrS_2$ have been simulated.[33, 34] An 18×18×1 (12×12×12) k-point grid is chosen to sample the Brillouin zone of the 2H and 1T phases of the monolayer (bulk) $CrS_2$ for the total energy calculations. On the other hand, an 18×12×1 (12×9×12) k-point grid is chosen to sample the Brillouin zone of the 1T' phases of the monolayer (bulk) $CrS_2$. We have used 100 Ry as the kinetic energy cut-off to expand the wavefunctions. The vacuum spacing in all monolayer calculations is 18 Å. All the structures are relaxed using DFT until the energy, forces, and pressures are less than $10^{-4}$ Ry, $10^{-3}$ Ry/Bohr, 0.5 kbar, respectively. The phonon band structures and frequencies at the Γ point are computed using density functional perturbation theory.[35] All the calculations are performed with Quantum ESPRESSO.

**Magnetism.** The superconducting quantum interference device (SQUID, Quantum Design) magnetometer is applied for the magnetic measurements. The applied magnetic field is 20 KOe and instrument sensitivity is $10^{-7}$ emu. Field-cooling magnetization is measured under a 20 KOe magnetic field. High temperature measurements are conducted with the oven module in SQUID using a heater stick.

**Magneto-optic Kerr effect (MOKE) measurements.** Polar magneto-optical Kerr effect (MOKE) measurements were performed in a 1K cryostat (CryoVac technologies) equipped with 9T out-plane and 1T in-plane magnetic field. A He-Ne laser of 633nm was linearly polarised using Glan-Thompson polariser and focused on the sample using a high numerical aperture (NA = 0.65) cryogenic objective with the spot size of ~1 μm. The laser was double modulated using a mechanical chopper at 2.3K Hz and photoelastic modulator (PEM, Hinds instruments) at 42.5K Hz. The retardation of PEM was fixed at 0.38 λ to produce alternating linearly, elliptically, and circularly polarised laser. The reflected laser beam was passed through a photoelastic modulator (PEM) and analyser (fixed at 45° with respect to polariser) which was collected by silicon photodetector with 40 dB gain and 4 mm active area. The Kerr rotation signal was extracted by lock-in detection using three Stanford lock-ins (SR830). The hysteresis loops were recorded by sweeping the magnetic field while focusing on the flake, the spatial scans were recorded by keeping the applied field constant and scanning the flake with cryogenic piezo scanners (Attocube).

**Conflicts of interest**

There are no conflicts to declare.

## Acknowledgments

C.M. would like to acknowledge the award of funding from the European Research Council (ERC) under the European Union's Horizon 2020 research and innovation programme (Grant Agreement No. 819069) and the award of a Royal Society University Research Fellowship (UF160539) and the Research Fellows Enhancement Award by the UK Royal Society (RGF/EA/180090) by the UK Royal Society UK, H.B. would like to acknowledge the support of the China Scholarship Council (CSC) Grant #201808060492. W.B. acknowledges EPSRC grant EP/X015661/1 for funding. SJH acknowledges funding from the ERC under the European Unions Horizon 2020 research and innovation programme (Grant ERC-2016-STG-EvoluTEM-715502. We also thank the Engineering and Physical Sciences Research Council (EPSRC) for funding under grants EP/Y024303, EP/S021531/1, EP/M010619/1, EP/V007033/1, EP/S030719/1, EP/V001914/1, EP/V036343/1 and EP/P009050/1 and also for the EPSRC Centre for Doctoral Training (CDT) Graphene-NOWNANO. TEM access was supported by the HenryRoyce Institute for Advanced Materials, funded through EPSRC grants EP/R00661X/1, EP/S019367/1, EP/P025021/1 and EP/P025498/1.